# *USABILITY TESTING* UNTUK MENGUKUR KEPUASAAN PENGGUNA *WEBSITE* MAILO STORE

**Andresia Pitri[1], Leon Andretti Abdillah[2]**
Fakultas Ilmu Komputer, Universitas Bina Darma
Email: ressypitri@gmail.com[1], leon.abdilah@yahoo.com[2]

***ABSTRACT***

*Technological developments in the current era of globalization cannot be avoided, with increasingly rapid progress to become technology as a medium of information that is very much needed in life. Mailo store is one of the SMEs that have used the website and the internet as a medium for storing their information to the wider community, therefore an evaluation is needed to measure the ease of website users for users using Usability Testing which has 8 variables, namely learning (learnability), Efficiency (efficiency), Easy to remember (memorability), Errors and security (errors), Easy to understand (Understandability), Easy to operate (Operability), attract user attention (Attractiveness), Satisfaction (satisfaction). This research data was collected by distributing questionnaires to 109 respondents using random sampling technique. Furthermore, the data obtained were processed using SPSS version 25 software. The data analysis method used was quantitative analysis method using validity and reliability tests, classical assumption tests, multiple regression tests and hypothesis testing. The results of the assessment of the usability testing method will be useful for the sustainability and development of the mailo store website in the future to make it even better.*

**ABSTRAK**

Perkembangan teknologi pada era globalisasi saat ini tidak dapt di hindari, dengan kemajuan yang semakin pesat menjadi teknologi sebagai media informasi yang begitu sangat dibutuhkan dalam kehidupan. Mailo store merupakan salah satu umkm yang telah menggunakan *website* dan *internet* sebagai media penyimpanan informasi mereka kepada masyarakat luas, maka dari itu sangat diperlukan evaluasi untuk mengukur kemudahan pengguna *website* bagi pemakainya menggunakan Pengujian Kebergunaan (*Usability Testing*) memiliki 8 variabel yaitu dipelajari (*learnability*), Efisiensi (*efficiency*), Mudah diingat (*memorability*), Kesalahan dan keamanan (*errors*), Mudah dipahami (*Understandability*), Mudah dioperasian (*Operability*), menarik perhatian pengguna (*Atractiveness*), Kepuasan (*satisfaction*). Data penelitian ini dikumpulkan dengan menyebarkan kuesioner kepada 109 responden dengan menggunakan teknik *sampel randon sampling*. Selanjutnya data yang diperoleh diolah menggunakan software SPSS versi 25. Metode analisa data yang digunakan adalah metode analisa kuantitatif dengan menggunakan uji validitas dan uji reabilitas, uji asumsi klasik, uji regresi berganda dan uji hipotesis. Hasil dari penilaian metode *usability testing* ini akan berguna untuk keberlangsungan dan perkembangan situs *website* mailo store kedepanya agar kebih baik lagi.

**Kata Kunci:** *Usability testing,* Kepuasan Pengguna*, Website*

## 1. PENDAHULUAN

Perkembangan teknologi informasi saat ini semakin meningkat apa lagi di era sekarang sebagai media informasi yang sangat di butuhkan dalam kehidupan sehari-hari. Sebagai media informasi tentu tidak bisa lepas dari *internet*, *Internet* adalah sebuah jaringan global yang menyatuhkan semua instalasi, perusahaan dan sebagainya. *Internet* yang telah tersebar luas di dunia seperti melalui situs *web*. Internet (Abdillah, 2020) memungkinkan pengiriman data dari sumber ke tujuan dalam berbagai bentuk.

Situs *web* adalah media untuk menyebarkan informasi dan promosi secara luas sebuah situs *website* dapat memeberikan keuntungan yang besar bagi pengguna karna dapat di akses dengan berbasis online penggunaan *website* saat ini berkembang dengan sangat cepat menjadikan salah satu alternatif media penyampaian informasi di segala bidang yang dapat digunakan tanpa ada batasan waktu dan tempat. Dan disamping itu, juga memungkinkan penggunanya untuk mendapatkan update data informasi dengan cepat di dapat kapan saja dan dimana saja (Aries, 2017). *Web* yang baik dapat memenuhi kepuasan penggunannya terhadap *web* tersebut maka untuk mengetahui apakah *web* sudah memenuhi kepuasan penggunanya, maka diperlukan suatu proses untuk mengukur dan terdapat beberapa metode yang dapat di gunakan, salah satunya adalah metode *Usabillity Testing*.

*Usabillity testing* adalah aribut kualitas yang menjelaskan atau mengukur seberapa mudah penggunaan suatu antar muka *(interface)*, Kata "*usabillity*" juga merujuk pada suatu metode untuk meningkatkan kemudahaan pemakaian selama proses desain (Nielsen, 2015) (ISO/IEC 9126). Untuk mengukur kepuasan pengguna menggunakan 8 variabel Kepuasan (*satisfaction*), 7 *independen* dan 1 *dependen* dalam penelitian ini yaitu sebagai berikut Kemudahan (*learnability*), Efisiensi (*efficiency*), Mudah diingat (*memorability*), Kesalahan dan keamanan (*errors*), Mudah dipahami (*Understandability*), Mudah dioperasian (*Operability*), Menarik perhatian pengguna (*Atractiveness*), Kepuasan (*satisfaction*).

Toko mailo store merupakan toko yang menjual berbagai macam fashion baju brand untuk remaja kekinian dan saat ini sudah memiliki *website* untuk pembelinnya yang bisa di akses melalui *link* https://mailostore.com/ yang di buat untuk mempermudah dalam mencari informasi dan mempermudah dalam hal pemesanan di mana saja dan kapan saja. Diharapkan hasil dari penilaian yang diperoleh dengan metode *usability* ini nanti menjadi sebuah tolak ukur dalam pengembangan situs *web* mailo store untuk kedepannya agar menjadi sebuah situs *web* yang lebih baik kedepannya sehingga bisa di jadikan acuan untuk kemajuan toko mailo store.

## 2. METODOLOGI PENELITIAN

### 2.1 Analisis Kuantitatif

Metode analisis kuantitatif dapat diartikan sebagai metode penelitian yang berlandaskan pada filsafat positivisme, digunakan untuk meneliti pada populasi atau sampel tertentu, pengumpulan data menggunakan instrumen penelitian, analisis data bersifat kuantitatif atau statistik dengan tujuan untuk menguji hipotesis yang telah ditetapkan (Sugiyono, 2017).

### 2.2 Analisis Statistik Deskriptif

Analisis statistik deskriptif adalah uji coba dengan menggunakan hasil seluruh isian kuesioner *website* mailo store dengan mencari tau *mean* dan *medium* dari jawaban responden. Hasil ini didapatkan dengan melakukan beberapa aktivitas perhitungan dengan menggunakan fitur dari aplikasi bantuan SPSS, diantaranya adalah dengan menggunakan fitur *Analyze, frequencies* dan *Descriptive Statistics* (Aprilian, 2014).

### 2.3 Pengujian Kualitas Data

Pengumpulan data yang dilakukan dalam penelitian ini adalah dengan menggunakan kuesioner, oleh karena itu untuk mengelola datanya maka dilakukan uji validitas dan uji reliabilitas yang berguna untuk menguji kesungguhan jawaban responden. Pengujian ini dilakukan dengan menggunakan aplikasi program SPSS *(Statistical Package for Social Sciences) for Windows Versi* 25.

1. Uji Validitas

Uji validitas merupakan derajat ketepatan antara data yang terjadi pada objek penelitian dengan daya yang dapat dilaporkan oleh peneliti. Dengan demikian data yang valid adalah data "yang tidak berbeda" antar data yang dilaporkan oleh peneliti dengan data yang sesungguhnya terjadi pada objek penelitian (Sugiyono, 2017).

Pengujian validitas data dalam penelitian ini dilakukan secara statistik yaitu menghitung korelasi antara masing-masing pernyataan dengan skor menggunakan metode product *moment pearson correlation.* Data dinyatakan valid jika nilai r hitung yang menggunakan nilai dari Corrected Item Total Correlation > dari r tabel pada signifikansi 0,05 (5%) (Ghozali, 2018).

2. Uji Reliabilitas

Reliabilitas adalah istilah yang dipakai untuk menunjukkan sejauh mana hasil pengukuran relatif konsisten apabila alat ukur digunakan berulang kali. Untuk menguji reliabilitas kuisioner digunakan Croanbach Alpa, reliabilitas suatu instrumen memiliki tingkat realiabilitas yang tinggi apabila nilai koefesien Croanbach Alpa yang dipeoleh >0,60% (Mz, 2019).

**2.4 Uji Hipotesis**

1. Uji F

Uji F digunakan untuk mengetahui hubungan antara variabel *independen* dan variabel *dependen*, apakah Kemudahan (*learnability*), Efisiensi (*efficiency*), Mudah diingat (*memorability*), Kesalahan dan keamanan (*errors*), Mudah dipahami (*Understandability*), Mudah dioperasian (*Operability*), Menarik perhatian pengguna (*Atractiveness*), Kepuasan (*satisfaction) website* mailo store. Langkah-langkah pengujiannya adalah sebagai berikut (Ghozali, 2014):

2. Uji T

Uji T pada penelitian ini digunakan untuk mengetahui apakah variabel Kemudahan (*learnability*), Efisiensi (*efficiency*), Mudah diingat (*memorability*), Kesalahan dan keamanan (*errors*), Mudah dipahami (*Understandability*), Mudah dioperasian (*Operability*), Menarik perhatian pengguna (*Atractiveness*), berprngaruh terhadap kepuasan pengguna *website* mailo store. Langkah-langkah pengujiannya adalah sebagai berikut (Ghozali, 2014)

**2.5 Uji Asumsi Klasik**

1. Uji Normalitas

Tujuan dilakukannya uji normalitas adalah untuk mengetahui apakah model regresi, variabel terikat dan variable bebas keduanya mempunyai distribusi normal atau tidak (Ghozali, 2018). Data yang berdistribusi normal dalam suatu model regresi dapat dilihat pada grafik normal P-P plot, dimana bila titik-titik yang menyebar disekitar garis diagonal serta penyebarannya mengikuti arah garis diagonal, maka data tersebut dapat dikatakan berdistribusi normal.

2. Uji Heteroskedastisitas

Uji Heteroskedastisitas bertujuan untuk menguji apakah dalam model regresi terjadi ketidaksamaan varians dari residual satu pengamatan ke pengamatan lain (Ghozali, 2018)
Dasar pengambilan keputusan adalah sebagi berikut. Jika ada data yang membentuk pola tertentu, seperti titik-titik yang membentuk pola tertentu dan teratur (bergelombang, melebar kemudian meyempit) dan Jika tidak ada pola yang jelas serta titik-titik menyebar diatas dan dibawah angka 0 pada sumbu Y, maka tidak terjadi heterokedastisitas.

3. Regresi Berganda

Evaluasi regresi pada dasarnya adalah studi mengenai ketergantunagn variabel dependen (terikat) dengan satu atau lebih variabel independen dengan tujuan untuk mengestimasi dan mempresiksi rata-rata populasi atau nilai-nilai variabel dependen berdasarkan nilai variabel independen yang diketahui (Latan, 2014).

## 2.6 Populasi dan Teknik Pengambilan Sampel

Populasi adalah sebagai suatu kumpulan subjek, *variabel*, konsep, atau fenomena. Kita dapat meneliti setiap anggota populasi untuk mengetahui sifat populasi yang bersagkutan ( Morison, 2012). Populasi dalam penelitian ini adalah pembeli melalui *website* mailo store dan pemilik sebanyak 149 orang pada tahun 2020.

Sampel adalah bagian dari jumlah dan karakteristik yang dimiliki oleh populasi tersebut (Sugiyono, 2017). Penelitian ini menggunakan teknik pengambilan sampel (Sampling) karena peneliti tidak mampu menjangkau keseluruhan populasi. Pada penelitian ini digunakan rumus slovin untuk menentukan sampel minimal dengan menggunakan taraf sgnifikan 5%.

$$n = \frac{N}{1 + N(e)^2}$$

$$n = \frac{149}{1 + 0{,}3725\ (0{,}05)^2}$$

$$n = \frac{145}{1 + 1{,}3725\ (0{,}0025)}$$

$$n = \frac{4154}{1{,}3725}$$

n = 108,56 → dibulatkan menjadi 109

Berdasarkan pendapat ini maka dalam penelitian ini sampel yang digunakan sebanyak 109 orang.

## 3. HASIL DAN PEMBAHASAN

### 3.1 Identitas Responden

Responden dalam penelitian ini adalah pengguna *website* Mailo Store dengan sampel sebanyak 109 responden yang terdiri dari 3 kategori yaitu berdasarkan jenis kelamin, usia, pekerja. Berikut ini akan dijelaskan identitas responden pengguna *website* Mailo Store.

### 3.2 Jenis Kelamin Responden

Dari hasil penyebaran kuesioner pada sampel sebanyak 109 responden berdasarkan jenis kelamin hasil penelitian ini. Berikut ini merupakan diagram presentase responden berdasarkan jenis kelamin:
1. Responden berjenis kelamin laki-laki sebanyak 65 orang (59,6%)
2. Responden berjenis kelamin Wanita sebanyak 44 orang (40,4%)

### 3.3 Umur Responden

Dari hasil penyebaran kuesioner pada sampel sebanyak 109 responden berdasarkan katagori usia responden hasil sebagai berikut

**Tabel 1 Umur Responden**

|  |  | Frequency | Percent | Valid Percent | Cumulative Percent |
|---|---|---|---|---|---|
| Valid | Generasi Z | 59 | 54.1 | 54.1 | 54.1 |
|  | Generasi Milenial | 40 | 36.7 | 36.7 | 90.8 |
|  | Generasi X | 10 | 9.2 | 9.2 | 100.0 |
|  | Total | 109 | 100.0 | 100.0 |  |

### 3.4 Pekerjaan Responden

Dari hasil penyebaran kuesioner pada sampel sebanyak 109 responden berdasarkan pekerjaan. Hasil penelitian menunjukkan terdapat 27,5% responden memiliki pekerjaan sebagai mahasiswa, 4,6% responden memiliki pekrjaan sebagai pegawai negeri, 34,9% responden memiliki pekerjaan sebagai pegawai swasta dan terdapat 33% responden memiliki pekerjaan sebagai lainnya.

### 3.5 Analisis Kuantitatif

Skor aktual adalah jawaban seluruh responden atas kuesioner yang telah diisi. Skor ideal adalah skor atau bobot tertinggi atau semua responden diasumsikan memilih jawaban dengan skor tertinggi.

$$\text{Skor Total} = \frac{\text{Skor Aktual}}{\text{Skor Ideal}} \times 100\%$$

$$\text{Skor Total} = \frac{10,288}{13,080} \times 100\% = 78\%$$

Pengukuran setiap variabel dilakukan secara terpisah untuk mengetahui skor total dari masing-masing variabel. Diketahui bahwa tingkat kepuasan pengguna terhadap *website* mailo store diperoleh nilai skor sebesar 78% yang berarti tanggapan pengguna setuju.

### 4. Analisis Statistik Deskriptif

Perhitungan *Mean* dan *Median* dalam analisis statistik deskriptif dimaksudkan untuk menemukan nilai rata-rata dari jawaban responden pada kuesioner.

**Tabel 2** Analisis Statistik Deskriptif

| No | Variabel | Indikator | Mean | Medium | Hasil |
|---|---|---|---|---|---|
| 1 | *Learnability* | Mudah dimengerti | 4,27 | 4,00 | Sangat Setuju |
|  |  | Mudah mencari informasi spesifik | 4,07 | 4,00 | Sangat Setuju |
|  |  | Mudah mengidentifikasi mekanisme navigasi | 3,89 | 4,00 | Setuju |

| No | Aspek | Indikator | Mean | Median | Keterangan |
|---|---|---|---|---|---|
| 2 | *Eficiency* | Mudah dijangkau dengan cepat | 4,25 | 4,00 | Sangat Setuju |
| | | | 3,89 | 4,00 | Setuju |
| | | Mudah dinavigasi | 3,70 | 4,00 | Setuju |
| 3 | *Memorability* | Mudah diingat | 4,28 | 4,00 | Sangat Setuju |
| | | | 4,28 | 4,00 | Sangat Setuju |
| | | Mudah dibangun kembali | 4,05 | 4,00 | Setuju |
| 4 | *Errors* | Sedikit jumlah kesalahan yang terdeteksi | 3,93 | 4,00 | Setuju |
| | | | 3,73 | 4,00 | Setuju |
| | | Kenyamanan untuk digunakan | 3,57 | 4,00 | Setuju |
| 5 | *Understandbility* | Mudah dipahamitanpa intruksi khusus | 4,06 | 4,00 | Sangat Setuju |
| | | Memamhami informasi yang disajikan | 3,95 | 4,00 | Setuju |
| | | Menu dan fitur mudah dipahami | 3,88 | 4,00 | Setuju |
| 6 | *Operability* | Mudah mengopersikan menu dan fitur | 3,92 | 4,00 | Setuju |
| | | | 3,91 | 4,00 | Setuju |
| | | Mudah mengoperasikan situs *web* | 3,69 | 4,00 | Setuju |
| 7 | *Atractiviness* | Tertarik untuk digunakan lagi | 4,20 | 4,00 | Sangat Setuju |
| | | | 3,97 | 4,00 | Setuju |
| | | Mmenarik untuk di rekomendasikan | 3,74 | 4,00 | Setuju |
| 8 | *Sastifaction* | Sistem yang menyenangkan untuk digunakan | 4,63 | 4,00 | Sangat Setuju |
| | | | 4,27 | 4,00 | Sangat Setuju |
| | | Kenyamanan untuk digunakan | 4,00 | 4,00 | Setuju |

## 5. Pengujian Kualitas Data

### 1. Uji Validitas

Seperti yang telah dijelaskan sebelumnya bahwa uji validitas digunakan untuk mengukur sah atau tidaknya suatu kuesioner. uji validitas dilakukan dengan membandingkan nilai r hitung lebih besar dari r tabel maka item tersebut dinyatakan valid. Dalam penelitian ini untuk *degree of freedom* (df) = n – 2, dalam hal ini n adalah jumlah responden yang berjumlah 109 responden, jadi df = 109-2 = 107. Dengan tingkat signifikan 0,05 maka didapat r tabel sebesar 0,1882.

**Tabel 3** Hasil Uji Validitas

| Variabel | r hitung | r table | Kondisi | Kesimpulan |
|---|---|---|---|---|
| **Kemudahan (*learnability*)** | | | | |
| X1_1 | 0,801 | 0,1882 | r hitung > r tabel | Valid |
| X1_2 | 0,859 | 0,1882 | r hitung > r tabel | Valid |
| X1_3 | 0,755 | 0,1882 | r hitung > r tabel | Valid |
| **Effisiensi (*efficiency*)** | | | | |
| X2_1 | 0,818 | 0,1882 | r hitung > r tabel | Valid |
| X2_2 | 0,854 | 0,1882 | r hitung > r tabel | Valid |
| X2_3 | 0,706 | 0,1882 | r hitung > r tabel | Valid |
| **Mudah diingat (*memorability*)** | | | | |
| X3_1 | 0,809 | 0,1882 | r hitung > r tabel | Valid |
| X3_2 | 0,728 | 0,1882 | r hitung > r tabel | Valid |
| X3_3 | 0,827 | 0,1882 | r hitung > r tabel | Valid |
| **Kesalahan dan Keamanan (*errors*)** | | | | |
| X4_1 | 0,707 | 0,1882 | r hitung > r tabel | Valid |
| X4_2 | 0,747 | 0,1882 | r hitung > r tabel | Valid |
| X4_3 | 0,804 | 0,1882 | r hitung > r tabel | Valid |
| **Mudah dipahami (*understandability*)** | | | | |
| X5_1 | 0,791 | 0,1882 | r hitung > r tabel | Valid |
| X5_2 | 0,861 | 0,1882 | r hitung > r tabel | Valid |
| X5_3 | 0,790 | 0,1882 | r hitung > r tabel | Valid |
| **Dioperasikan (*operability*)** | | | | |
| X6_1 | 0,829 | 0,1882 | r hitung > r tabel | Valid |
| X6_2 | 0,879 | 0,1882 | r hitung > r tabel | Valid |
| X6_3 | 0,795 | 0,1882 | r hitung > r tabel | Valid |
| **Menarik Perhatian Pengguna (*atractiviness*)** | | | | |
| X7_1 | 0,628 | 0,1882 | r hitung > r tabel | Valid |
| X7_2 | 0,793 | 0,1882 | r hitung > r tabel | Valid |
| X7_3 | 0,323 | 0,1882 | r hitung > r tabel | Valid |
| **Kepuasan (*satisfaction*)** | | | | |
| Y1 | 0,701 | 0,1882 | r hitung > r tabel | Valid |
| Y2 | 0,869 | 0,1882 | r hitung > r tabel | Valid |
| Y3 | 0,733 | 0,1882 | r hitung > r tabel | Valid |

2. Uji Reliabilitas

Reliabilitas dilakukan untuk mengetahui konsistensi alat ukur dalam mengukur gejala yang sama. Syarat untuk menyatakan jika item itu reliabel adalah dengan melihat hasil uji reliabilitas jika setiap variabel > dari 0,60 berarti variabel tersebut reliabel.

**Tabel 4** Uji Reliabilitas

| Variabel | *Cronbach's Alpha* | Keterangan |
|---|---|---|
| Kemudahan | 0,743 | Reliabel |
| Effisiensi | 0,691 | Reliabel |
| Mudah diingat | 0,722 | Reliabel |
| Kesalahan dan Keamanan | 0,619 | Reliabel |
| Mudah dipahamai | 0,746 | Reliabel |
| Dioperasikan | 0,779 | Reliabel |
| Menarik Perhatian Pengguna | 0,644 | Reliabel |
| Kepuasan | 0,653 | Reliabel |

6. **Uji Hipotesis**

1. Uji F

Untuk mengetahui tingkat signifikan pengaruh variabel-variabel independent secara bersama-sama simultan terhadap variabel dependen dilakukan dengan menggunakan uji F yaitu dengan cara membandingkan antara F hitung dengan F tabel.

**Tabel 5** Hasil Uji F

| | ANOVAª | | | | | |
|---|---|---|---|---|---|---|
| Model | | Sum of Squares | Df | Mean Square | F | Sig. |
| 1 | Regression | 116.245 | 7 | 16.606 | 21.059 | .000ᵇ |
| | Residual | 79.645 | 101 | .789 | | |
| | Total | 195.890 | 108 | | | |
| a. Dependent Variable: satisfaction | | | | | | |
| b. Predictors: (Constant), Atractiveness, errors, efficiency, Understandability, Operability, learnability, memorability | | | | | | |

Berdasarkan tabel 5 Output Regression ANOVA diketahui nilai F hitung sebesar 21,059 dengan nilai signifikan 0,000. Untuk F tabel dapat dicari dengan melihat pada tabel f dengan signifikan 0,05 dan menentukan df1 = k-1 atau 7-1 = 6, dan df2 = n-k atau 109-7 = 102 (n = jumlah data; k=jumlah variabel independen). Di dapat F tabel adalah sebesar 2,188. Jika apabila F hitung < F Tabel maka Ho diterima dan apabila F hitung >F Tabel maka Ho ditolak. Dapat diketahui bahwa F hitung (21,059) > F tabel (2,188) maka Ho ditolak. Jadi kesimpulannya yaitu kemudahan (*leabillity*), efisiensi (*efficiency*), mudah diingat (*memorability*), kesalahan atau keamanan (*errors*), mudah dipahami (*understandability*), dioperasikan (*operabillity*), menarik perhatian pengguna (*atractiviness*) secara bersama-sama berpengaruh terhadap kepuasan (*Satisfaction*).

2. Uji T

Uji T digunakan untuk menguji pengaruh variabel independen secara persial terhadap variabel dependen. Taraf signifikan yang ditentukan adalah menggunakan nilai 0,05. Berikut adalah perhitungan uji t dari tiap variabel independen.

**Tabel 6** Hasil Uji T

| | Coefficientsª | | | | | | |
|---|---|---|---|---|---|---|---|
| Model | | Unstandardized Coefficients | | Standardized Coefficients | T | Sig. | Collinearity Statistics | |
| | | B | Std. Error | Beta | | | Tolerance | VIF |
| 1 | (Constant) | 1.539 | .956 | | 1.611 | .110 | | |
| | Learnability | .172 | .056 | .227 | 3.091 | .003 | .749 | 1.335 |
| | Efficiency | .215 | .053 | .274 | 4.025 | .000 | .869 | 1.151 |
| | Memorability | .138 | .047 | .217 | 2.954 | .004 | .745 | 1.342 |
| | Errors | .046 | .049 | .067 | .949 | .345 | .802 | 1.247 |
| | Understandability | .147 | .054 | .192 | 2.744 | .007 | .820 | 1.219 |
| | Operability | .100 | .055 | .136 | 1.818 | .072 | .715 | 1.399 |
| *1.* | Atractiveness | .130 | .055 | .172 | 2.366 | .020 | .759 | 1.317 |
| | a. Dependent Variable: satisfaction | | | | | | | |

*Learnability* (X1)

Diketahui t hitung dari *learnability* adalah 3,091 (pada tabel 4.15 *Output Regression Coefficients*), t tabel dapat dicari pada tabel statistik pada signifikan 0,05/2=0,025(uji 2 sisi) dengan df = n-k-i atau 109-7-1=101 (k adalah jumlah variabel independen). Di dapat t tabel sebesar 1,984. Kesimpulan yang dapat diambil apabila t hitung < t tabel atau t hitung > t tabel jadi Ho diterima. Apabila t hitung > t tabel atau t hitung < t tabel jadi Ho ditolak. Dapat diketahui bahwa t hitung (3,091) > t tabel (1,984) jadi Ho ditolak, kesimpulan yaitu *Learnability* (X1) berpengaruh terhadap kepuasan pengguna (Y).

2. *Efficiency* (X2)

Diketahui t hitung dari *efficiency* adalah 4.025 (pada tabel 4.15 *Output Regression Coefficients*), t tabel dapat dicari pada tabel statistik pada signifikan 0,05/2=0,025(uji 2 sisi) dengan df = n-k-i atau 109-7-1=101 (k adalah jumlah variabel independen). Di dapat t tabel sebesar 1,984. Kesimpulan yang dapat diambil apabila t hitung < t tabel atau t hitung > t tabel jadi Ho diterima. Apabila t hitung > t tabel atau t hitung < t tabel jadi Ho ditolak. Dapat diketahui bahwa t hitung (4.025) > t tabel (1,984) jadi Ho ditolak, kesimpulan yaitu *Efficiency* (X2) berpengaruh terhadap kepuasan pengguna (Y).

3. *Memorability* (X3)

Diketahui t hitung dari *memorability* adalah 2.954 (pada tabel 4.15 *Output Regression Coefficients*), t tabel dapat dicari pada tabel statistik pada signifikan 0,05/2=0,025(uji 2 sisi) dengan df = n-k-i atau 109-7-1=101 (k adalah jumlah variabel independen). Di dapat t tabel sebesar 1,984. Kesimpulan yang dapat diambil apabila t hitung < t tabel atau t hitung > t tabel jadi Ho diterima. Apabila t hitung > t tabel atau t hitung < t tabel jadi Ho ditolak. Dapat diketahui bahwa t hitung (2.954) < t tabel (1,984) jadi Ho ditolak, kesimpulan yaitu *Memorability* (X3) berpengaruh terhadap kepuasan pengguna (Y).

4. *Errors* (X4)

Diketahui t hitung dari *errors of use* adalah 0,949 (pada tabel 4.15 *Output Regression Coefficients*), t tabel dapat dicari pada tabel statistik pada signifikan 0,05/2=0,025(uji 2 sisi) dengan df = n-k-i atau 109-7-1=101 (k adalah jumlah variabel independen). Di dapat t tabel sebesar 1,984. Kesimpulan yang dapat diambil apabila t hitung < t tabel atau t hitung > t tabel jadi Ho diterima. Apabila t hitung > t tabel atau t hitung < t tabel jadi Ho ditolak. Dapat diketahui bahwa t hitung (0,949) > t tabel (1,984) jadi Ho diterima, kesimpulan yaitu *Errors* (X4) tidak berpengaruh terhadap kepuasan pengguna (Y).

5. *Understandability* (X5)

Diketahui t hitung dari *understandability* adalah 2,744 (pada tabel 4.15 *Output Regression Coefficients*), t tabel dapat dicari pada tabel statistik pada signifikan 0,05/2=0,025(uji 2 sisi) dengan df = n-k-i atau 109-7-1=101 (k adalah jumlah variabel independen). Di dapat t tabel sebesar 1,984. Kesimpulan yang dapat diambil apabila t hitung < t tabel atau t hitung > t tabel jadi Ho diterima. Apabila t hitung > t tabel atau t hitung < t tabel jadi Ho ditolak. Dapat diketahui bahwa t hitung (2,744) > t tabel (1,984) jadi Ho ditolak, kesimpulan yaitu *Understandability* (X2) berpengaruh terhadap kepuasan pengguna (Y).

6. *Operabiity* (X6)

Diketahui t hitung dari *operabiity* adalah 1.818 (pada tabel 4.15 *Output Regression Coefficients*), t tabel dapat dicari pada tabel statistik pada signifikan 0,05/2=0,025(uji 2 sisi) dengan df = n-k-i atau 109-7-1=101 (k adalah jumlah variabel independen). Di dapat t tabel sebesar 1,984.

Kesimpulan yang dapat diambil apabila t hitung < t tabel atau t hitung > t tabel jadi Ho diterima. Apabila t hitung > t tabel atau t hitung < t tabel jadi Ho ditolak. Dapat diketahui bahwa t hitung (1.818) < t tabel (1,984) jadi Ho diterima, kesimpulan yaitu *Operabiity* (X6) tidak berpengaruh terhadap kepuasan pengguna (Y).

*7. Atractiviness* (X7)

Diketahui t hitung dari *activiness* adalah 2,366 (pada tabel 4.15 *Output Regression Coefficients*), t tabel dapat dicari pada tabel statistik pada signifikan 0,05/2=0,025 (uji 2 sisi) dengan df = n-k-i atau 109-7-1=101 (k adalah jumlah variabel independen). Di dapat t tabel sebesar 1,984. Kesimpulan yang dapat diambil apabila t hitung < t tabel atau t hitung > t tabel jadi Ho diterima. Apabila t hitung > t tabel atau t hitung < t tabel jadi Ho ditolak. Dapat diketahui bahwa t hitung (2,366) > t tabel (1,984) jadi Ho ditolak, kesimpulan yaitu *Atractiviness* (X7) berpengaruh terhadap kepuasan pengguna (Y).

**8. Uji Asumsi Klasik**

1. Uji Normalitas

Uji normalitas adalah pengujian tentang kenormalan distribusi data. Dari grafik terlihat bahwa nilai *plot* P-P terletak disekitar garis diagonal, *plot* P-P tidak menyimpang jauh dari garis diagonal sehingga dapat diartikan bahwa distribusi data normal regresi dapat dilihat pada gambar 1

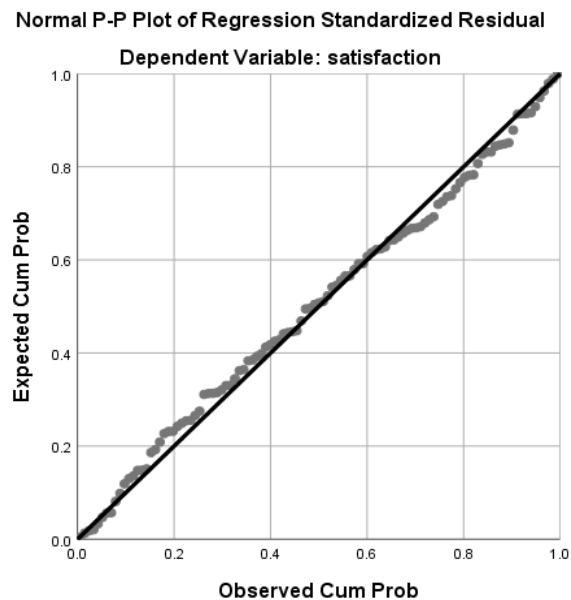

**Gambar 1** Diagram Grafik P-P Plot

2. Uji Heteroskedastisitas

Uji heterokedastisitas bertujuan untuk menguji apakah data model regresi terjadi ketidaksamaan *variance* dari suatu pengamatan kepenngamatan yang lain. Jika *variance* dari residual suatu pengamatan lain tetap, maka disebut heterokedastis. Cara untuk mendeteksi ada atau tidaknya heterokedastis adalah melihat grafik plot antara linai predeksi variable dependen *zpred* dengan residualnya *sresid*, dasar pengambilan keputusannya adalah sebagai berikut. Jika ada pola tertentu, seperti titik-titik yang ada membentuk pola tertentu yang teratur (bergelombang, melebar kemudian menyempit), maka mengindikasikan telah terjadi heteroskedastisitas. Jika tidak ada pola yang jelas, serta titik-titik menyebar diatas dan dibawah angka 0 pada sumbu Y, maka tidak terjadi heteroskedatisitas.

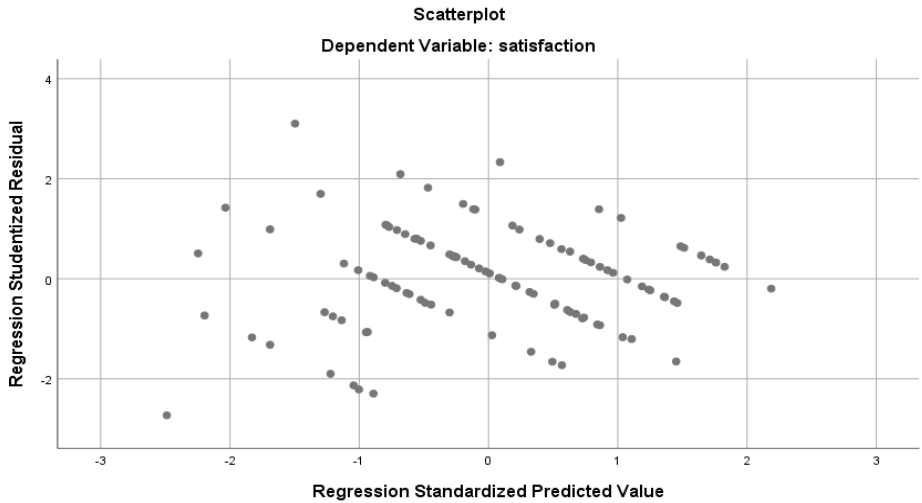

**Gambar 2** Uji Heteroskedastisitas

3. Hasil Regresi Berganda

Berdasarkan perhitungan regresi antara variabel *learnability, efficiency*, *memorability*, *errors*, *Understandability, Operability*, *Atractiveness*, *sastifaction* dengan menggunakan program SPSS 25, diperoleh hasil sebagai berikut.

**Tabel 7** Hasil Persamaan Berganda

| Model | R | R Square | Adjusted R Square | Std. Error of the Estimate |
|---|---|---|---|---|
| 1 | .770a | .593 | .565 | .888 |

Dari hasil diatas diketahui nilai Koefisien determinasi (regerial berganda) sebesar 0.770 menunjukkan bahwa pada variabel independen yaitu Kemudahan (*learnability*), Efisiensi (*efficiency*), Mudah diingat (*memorability*), Kesalahan dan keamanan (*errors*), Mudah dipahami (*Understandability*), Mudah dioperasian (*Operability*), menarik perhatian pengguna (*Atractiveness*) dan kepuasan (*sastifaction*) memberikan pengaruh sebesar 77% sedangkan sisanya sebesar 23% dipengaruhi oleh faktor lain.

**4. KESIMPULAN**

Berdasarkan hasil dan pembahasan diatas, maka dapat dibuat beberapa kesimpilan bahwa penelitian ini bertujuan untuk mengetahui pengaruh instrument *Usabillity Testing* terdiri dari variabel Kemudahan (*learnability*), Efisiensi (*efficiency*), Mudah diingat (*memorability*), Kesalahan dan keamanan (*errors*), Mudah dipahami (*Understandability*), Mudah dioperasian (*Operability*), menarik perhatian pengguna (*Atractiveness*) terhadap kepuasan penggunaan *website* Mailo Store penelitian ini menggunakan data primer yang diperoleh dari kuesioner yang menggunakan pengukuran skala likert. Kuesioner dibagikan kepada responden yang menggunakan *website* mailo store dan kuesioner yang dibagikan adalah sebanyak 109 kuesioner. Data kuesioner yang sudah terkumpul akan diolah menggunakan bantuan *software* SPSS 25 *for windows.*

Pertanyaan yang terdapat dalam kuesioner dilakukan uji validitas dan uji reabilitas. Dari uji validitas dapat disimpulkan bahwa pertanyaan yang terdapat dalam kuesioner bernilai valid karena nilai signifikansi dari masing-masing indikator variabel besarnya kurang dari 0,05. Sedangkan uji reabilitas digunakan untuk mengukur suatu kuesioner yang merupakan indikator dari variabel atau reliabel jika memberikan nilai *Cronbach-Alph(a)* dari masing-masing variabel nilainya lebih besar dari 0,60.

Penelitian ini digambarkan dalam model regresi berganda yaitu menganalisis pengaruh instrument *Usabillity Testing* terdiri dari variabel Kemudahan (*learnability*), Efisiensi (*efficiency*), Mudah diingat (*memorability*), Kesalahan dan keamanan (*errors*), Mudah dipahami (*Understandability*), Mudah dioperasian (*Operability*), menarik perhatian pengguna (*Atractiveness*) terhadap kepuasan penggunaan *website* mailo store. Hasil penelitian menunjukan bahwa tidak semua factor yang bergabung dalam instrument *Usabillity Testing* berpengaruh terhadap kepuasan pengguna *website* mailo store. Dari tujuh faktor yang tergabung dalam instrument *Usabillity Testing* pada *website* mailo store hanya Kemudahan (*learnability*), Efisiensi (*efficiency*), Mudah diingat (*memorability*), Mudah dipahami (*Understandability*), menarik perhatian pengguna (*Atractiveness*) yang menunjukan pengaruh signifikan terhadap kepuasan pengguna *website* mailo store. Sedangkan factor yang terdapat dalam instrument *Usabillity Testing* seperti Kesalahan dan keamanan (*errors*), dan Mudah dioperasian (*Operability*) tidak mennjukan adanya pengaruh signifikan terhadap kepuasaan pengguna *wwebsite* mailo store.

**DAFTAR PUSTAKA**